\documentclass[conference]{IEEEtran}
\usepackage{graphicx,amsmath,amssymb}
\usepackage{subfigure}
\usepackage{citesort}
\usepackage{fancyhdr}
\usepackage{mdwmath}
\usepackage{mdwtab}
\usepackage{balance}
\usepackage{xcolor}
\usepackage{bm}
\usepackage{bbm}
\usepackage{stfloats}
\usepackage{cite}
\usepackage{epstopdf}
\usepackage{algorithm}
\usepackage{algorithmic}
\usepackage{multirow}
\usepackage{flafter}

\usepackage{amssymb}
\usepackage{amsmath}

\usepackage[
top    = 0.7 in,
bottom = 1.05 in,
left   = 0.63 in,
right  = 0.63 in]{geometry}

\newtheorem{theorem}{Theorem}

\newtheorem{lemma}{Lemma}

\newtheorem{corollary}{Corollary}

\begin{document}
\title{Downlink Analysis for Reconfigurable Intelligent Surfaces Aided NOMA Networks}

\author{
\IEEEauthorblockN{Chao~Zhang\IEEEauthorrefmark{2}, Wenqiang~Yi\IEEEauthorrefmark{2}, Yuanwei~Liu\IEEEauthorrefmark{2}, Zhijin~Qin\IEEEauthorrefmark{2}, and Kok~Keong~Chai\IEEEauthorrefmark{2}} \IEEEauthorblockA{\IEEEauthorrefmark{2}Queen Mary University of London, London, UK} }
\maketitle

\begin{abstract}
  By activating blocked users and altering successive interference cancellation (SIC) sequences, reconfigurable intelligent surfaces (RISs) become promising for enhancing non-orthogonal multiple access (NOMA) systems. This work investigates the downlink performance of RIS-aided NOMA networks via stochastic geometry. We first introduce the unique path loss model for RIS reflecting channels. Then, we evaluate the angle distributions based on a Poisson cluster process (PCP) framework, which theoretically demonstrates that the angles of incidence and reflection are uniformly distributed. Lastly, we derive closed-form expressions for coverage probabilities of the paired NOMA users. Our results show that 1) RIS-aided NOMA networks perform better than the traditional NOMA networks; and 2) the SIC order in NOMA systems can be altered since RISs are able to change the channel gains of NOMA users.
\end{abstract}

\section{Introduction}
Due to introducing new freedom, non-orthogonal multiple access (NOMA) evolves into a promising technique. By sharing spectrum with power multiplexing schemes and successive interference cancellation (SIC), the spectral efficiency and user connectivity can be significantly improved to satisfy different target requirements~\cite{YuanweiNOMA}. In spite of benefits, NOMA techniques still have several important implementation challenges such as lower received power and severer interference than orthogonal multiple access (OMA) users. To cope with the challenges, reconfigurable intelligent surfaces (RIS) hold promise in several aspects. On the one hand, environmental obstacles may ruin the required channel condition, especially for the far NOMA users blocked by high buildings, which results in inevitable outage situations. With the aid of RISs under line-of-sight (LOS) propagation, we are able to exploit reflecting links through RISs to improve the channel condition of blocked NOMA users~\cite{8910627}. On the other hand, RISs are enabled to adjust the SIC order in NOMA networks. Note that channel state information (CSI) determines SIC sequences, which means the NOMA user with weaker channel gain than others is decoded preferentially. Adding reflecting links via RISs enhances the channel conditions, thereby the rank of statistical CSI is altered contributing to different decoding orders. Based on this method, we are capable of avoiding primary NOMA users experiencing redundant SIC procedures that will increase the outage probability of these users.

A RIS can be regarded as a two-dimensional-equivalent reconfigurable meta-material, which is consist of elementary elements called scattering particles or meta-atoms \cite{RISURSorRIS}. Based on intelligent meta-surface technologies, RISs have properties such as absorbing incident waves or modifying the reflected wavefronts. In contrast to mirrors, RISs are able to adjust the angle of reflection and electric field strength. A major and basic open research challenge is investigating the path loss model of RIS reflecting channels. Recent research contributions have studied the path loss model based on two typical methods, which are 1) correlated to the sum of incidence and reflection distances; and 2) correlated to the product of these distances. According to a fundamental work \cite{RISURSorRIS}, both of the typical methods are correct but utilizing in different application scenarios such that: 1) ``sum of distances'' model is suitable for short-distance communications such as indoor scenarios, while 2) ``product of distances'' model is suitable for long-distance communications such as outdoor scenarios. To reduce the path loss and interference, RISs are placed near to the served NOMA user. This spatial grouping property can be depicted by a tractable stochastic geometry model, namely Poisson cluster process (PCP)~\cite{7982794,8856258}, which provides a theoretical framework for investigating the average performance of RIS-aided NOMA networks. Therefore, this paper introduces a general case of path loss models, followed by an application of the ``product of distances'' model to analyze multi-cell NOMA networks for outdoor scenarios via a PCP-based spatial model.

Motivated by 1) enhancing the channel quality of blocked users; and 2) altering the SIC order to keep primary NOMA users avoiding SIC procedures, we investigate RIS-aided multi-cell NOMA networks. The main contributions are summarized as follows: 1) based on \cite{RISURSorRIS}, we derive the path loss model of RIS reflecting links in long-distance regions; 2) by modeling the multi-cell networks as a PCP distribution, we investigate the angle distributions, which verify that the angles formed by users, RISs and base stations (BSs) are uniformly distributed in $[0,\pi]$; 3) we derive closed-form expressions for coverage probabilities of the paired NOMA users to enhance the evaluation efficiency; and 4) numerical results illustrate that RIS-aided NOMA networks acquire significantly enhanced performance than traditional NOMA scenarios.

\section{System Model}

This paper considers RIS-aided downlink NOMA networks, where BSs and users are modeled according to two independent homogeneous Poisson point process (HPPPs), namely $\Phi_b\subset \mathbb{R}^2$ with density $\lambda_b$ and $\Phi_u\subset \mathbb{R}^2$ with density $\lambda_u$, respectively. Two-user NOMA clusters are served by orthogonal frequencies to cancel the intra-cell interference. We assume that one of the paired users has already been connected to a BS in the previous user association process~\cite{7982794}. The other one, namely the typical user, joins this occupied resource block by applying power-domain NOMA techniques. To simplify the analysis, the connected user is not included in the user set $\Phi_u$ and the distance between this user to its BS is invariable as $r_c$. The typical user is randomly selected from $\Phi_u$ and its location is fixed at the origin of the considered plane.

\subsection{RIS-aided Link Model}
RIS-aided networks as one promising application of RISs are to enhance the performance of blocked users by providing LoS transmission. We assume one RIS is employed for helping the typical user. Based on stochastic geometry methods, the typical user network is modeled through the Matern cluster processes (MCP) pattern of PCP models with fixed nodes in clusters. Hence, there are three different links: 1) BU link, the link between the typical user and one BS; 2) BR link, the link between the BS and the RIS; and 3) RU link, the link between the RIS and the typical user.

This work focuses on analyzing a blocked typical user and the RIS is applied to establish LoS route between the typical user and BSs~\cite{8910627}. Therefore, the BU link is assumed to be NLoS and the BR and RU links are LoS. Moreover, all NLoS communications are ignored in this paper due to their negligible received power.

To ensure the RU link is LoS, the RIS is uniform distributed in a ball area of the typical user with radius $R_L$, denoted as $\mathbb{O}(0,R_L) \subset \mathbb{R}^2$, where $\mathbb{O}(a,b)$ represents an annulus with the inner radius $a$ and outer radius $b$. Regarding the BR link, since large RISs are commonly deployed at a high height, we assume the BR link is always LoS. Due to considering a blocked typical user, the region of considered BSs is in the range $\mathbb{O}(R_L,\infty)$\footnote{For BSs located in the RIS ball area $\mathbb{O}(0,R_L)$, RISs may weaken their direct LoS transmission due to phase difference~\cite{8910627}. Coherent transmission is desired for this case, which is beyond the scope of this paper. We will study it in the future.}. The association criterion for the typical user is to associate with the BS with the highest received power, which equals that the distance between the RIS and the associated BS is the nearest. We assume the locations of the RIS and the associated BS are $\mathbf{x}_R$ and $\mathbf{x}_B$. Therefore, the distance between the associated BS and the RIS is correspondingly expressed as
 \begin{align}
 \mathbf{x}_{BR} = \arg \min_{\mathbf{x}_B \in \Phi_{B}} \|\mathbf{x}_B-\mathbf{x}_R \|
 \end{align}
 where $\mathbf{x}_R \in \mathbb{O}(0,R_L)$, $\Phi_{B} \subset\Phi_b$, $\Phi_{B} \subset \mathbb{O}(R_L,\infty)$ and the location of arbitrary interfering BS can be denoted by $\mathbf{x}_I \in \Phi_{r} \setminus  \mathbf{x}_B$.

 \begin{figure*}[t]
 \vspace{-0.3cm}
\centering
\includegraphics[width= 5in]{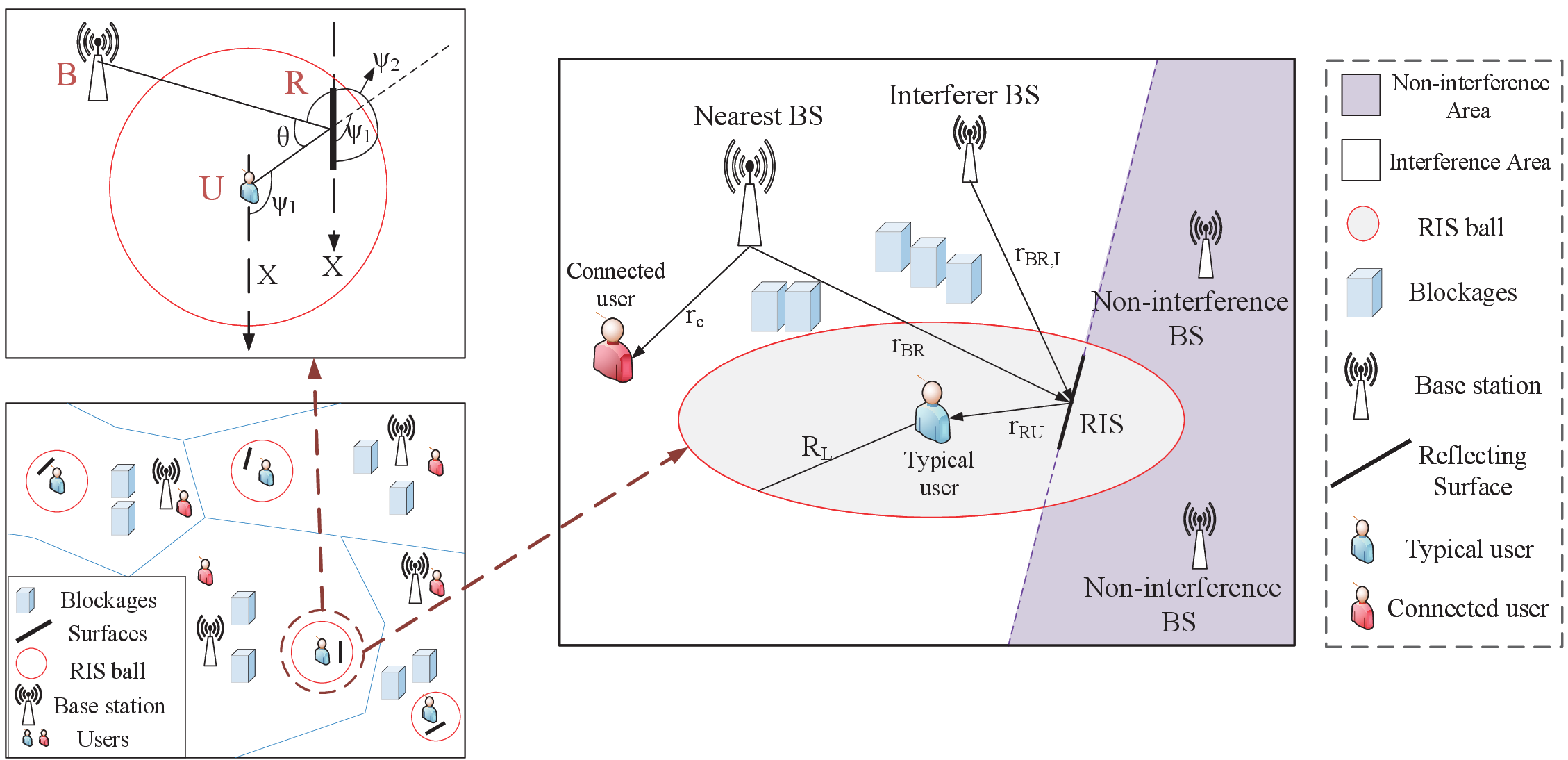}
\caption{(a) Bottom-left: An illustration of RIS-aided multi-cell scenarios; (b) Top-left: A typical NOMA network with angle demonstrations; (c) Wright: A typical NOMA network with two types of BSs, such as interferer BSs facing RISs and Non-interference BSs against RISs.}
\vspace{-0.4cm}
\label{fig_1}
\end{figure*}

\subsection{Path Loss Model}

 This work considers a $2L$ linear RIS, whose central is located at $\mathbf{x}_R=\left(x_R,y_R\right)\subset \mathbb{O}(0,R_L)$ and angle to the x-axis is $\theta_R$. We assume the associated BS and the typical user are distributed in the same side of the RIS to establish reflecting transmission~\cite{di2019reflection}. Therefore, the coordinate of one point on the RIS with a distance $x\in[-L,L]$ can be expressed as $\mathbf{x}_R(x)=\left(x_R+x\cos(\theta_R),y_R+x\sin(\theta_R)\right)\subset \mathbb{R}^2$, where $\theta_R \in \left[\min\{\angle \mathbf{x}_R, \angle (\mathbf{x}_B-\mathbf{x}_R)\},\max\{\angle \mathbf{x}_R, \angle (\mathbf{x}_B-\mathbf{x}_R)\}\right]$.

 Under a high-frequency case with cylindrical electromagnetic (EM) waves~\cite{ntontin2019reconfigurable}, if one transmitting BS is located at $\mathbf{x}_b\in\{\mathbf{x}_B, \mathbf{x}_I\}$, the path loss model for the typical user is given by~\cite{RISURSorRIS}
 \begin{align}\label{P_t}
 P_t(\mathbf{x}_b,\mathbf{x}_R) = \left|\int_{ - L}^{ + L} {\Psi\left( x \right)} \exp \left( { - jk\Omega\left( x \right)} \right)dx\right|^2,
 \end{align}
 where
 \begin{align}
 &\Psi\left( x \right) = \frac{ {\cos \left( {{\theta _{\mathrm{BR}}(x)}} \right) + \cos \left( {{\theta _{\mathrm{RU}}(x)}} \right)} }{{8\pi {\sqrt{ {{r_{\mathrm{BR}}(x)}{r_{\mathrm{RU}}(x)}} }}}},\\
 &\Omega\left( x \right) = {r_{\mathrm{BR}}}\left( x \right) + {r_{\mathrm{RU}}}\left( x \right) - \Theta \left( x \right).
\end{align}
  The communication distance for the BR and the RU links are $r_{\mathrm{BR}} = \|\mathbf{x}_b-\mathbf{x}_R(x)\|$ and $r_{\mathrm{RU}} = \|\mathbf{x}_R(x)\|$, respectively. Considering the reflecting point is at $\mathbf{x}_R(x)$, the the angle of incidence $\theta _{\mathrm{BR}}(x)\in\left(0,\frac{\pi}{2}\right]$ represents the angle between the corresponding BR link and the perpendicular bisector of the RIS, whilst the angle of reflection $\theta _{\mathrm{RU}}(x)\in\left(0,\frac{\pi}{2}\right]$ is the angle between the corresponding RU link and the perpendicular bisector of the RIS. The $\Theta \left( x \right)$ is the phase-shifting parameter of RISs which is decided by the desired transmitter and receiver.

 For the connected user at $\mathbf{x}_c \subset \mathbb{R}^2$, we assume the transmission between this user and its BS follows traditional wireless communications. Therefore, the path loss expression of the connected user is as follows
 \begin{align}
  P_c(\mathbf{x}_b,\mathbf{x}_c) = C\|\mathbf{x}_b-\mathbf{x}_c\|^{-\alpha_c},
 \end{align}
 where the $C$ is the intercept and $\alpha_c$ is the path loss exponent for the direct link. Note that the distance between the connected user and the associated BS is fixed. Therefore, $r_c = \|\mathbf{x}_B-\mathbf{x}_c\|$ is a constant.

 \subsection{Signal Model}
 In order to guarantee the quality of service (QoS) of the connected user, we assume the associated BS allocates more transmit power to the connected user than the typical user and SIC is processed at the typical user. Therefore, the signal-to-interference-plus-noise-ratio (SINR) for the SIC process at the typical user is given by
  \begin{align}
{\gamma _\mathrm{SIC}} = \frac{{{a_c}{P_b}{{\left| {{h_{t,\mathbf{x}_B}}} \right|}^2}P_c(\mathbf{x}_B,\mathbf{x}_c)}}{{ {{a_t}{P_b}{{\left| {{h_{t,\mathbf{x}_B}}} \right|}^2}P_t(\mathbf{x}_B,\mathbf{x}_R)} +I_{t,\rho_t} + {\sigma ^2}}},
 \end{align}
 where
 \begin{align}
 I_{t,\rho_t}=\rho_t\sum\limits_{\mathbf{x}_I \in \Phi_{r} \setminus  \mathbf{x}_B} {{P_b}{{\left| {{h_{t,\mathbf{x}_I}}} \right|}^2}P_t(\mathbf{x}_I,\mathbf{x}_R)}
 \end{align}
 and $P_b$ is the transmit power of BSs in each NOMA cluster and $\sigma ^2$ is the variance of additive white Gaussian noise (AWGN). $a_t$ and $a_c$ are the power allocation parameters for the typical user and the connected user, respectively. Moreover, $a_c > a_t$ and $a_c + a_t = 1$. When the transmitter is at $\mathbf{x}$, for the receiver $\kappa$, the ${h_{\kappa,\mathbf{x}}}$ represents its Nakagami fading term with an integer parameter $m_\kappa$~\cite{8876629}. Additionally, $\kappa = c$ means the receiver is the connected user and $\kappa = t$ means that is the typical user. Regarding the interference $I_{t,\rho_t}$, since the signal from the back of RISs cannot pass through RISs, we assume $\rho_t \in [0,1]$ of the entire interference is able to reach the receiver $\kappa$.

 After the SIC process, the typical user decodes its data. The decoding SINR can be expressed as
\begin{align}
{\gamma _t} = \frac{{{a_t}{P_b}{{\left| {{h_{t,\mathbf{x}_B}}} \right|}^2}P_t(\mathbf{x}_B,\mathbf{x}_R)} }{{I_{t,\rho_t} + {\sigma ^2}}}.
\end{align}

For the connected user, it directly decodes its messages by regarding the partner's signal as interference. Therefore, the decoding SINR for the connected user is as follows
\begin{align}
{\gamma _c} = \frac{{{a_c}{P_b}{{\left| {{h_{c,\mathbf{x}_B}}} \right|}^2}P_c(\mathbf{x}_B,\mathbf{x}_c)}}{{ {{a_t}{P_b}{{\left| {{h_{c,\mathbf{x}_B}}} \right|}^2}P_t(\mathbf{x}_B,\mathbf{x}_R)} +I_{c} + {\sigma ^2}}},
 \end{align}
 where
  \begin{align}
 I_{c}=\sum\limits_{\mathbf{x}_I \in \Phi_{r} \setminus  \mathbf{x}_B} {{P_b}{{\left| {{h_{t,\mathbf{x}_I}}} \right|}^2}P_c(\mathbf{x}_I,\mathbf{x}_c)}.
 \end{align}

It is worth noting that the connected user can be interfered by all BSs excepting the associated BS.

\section{Path loss Analysis}

In this section, several notifications are outlined without loss of generality that: 1) all the BSs are modeled as point sources; 2) surfaces are deployed based on long-distance communication models, thereby the surfaces receive directional lights with approximations as ${r_Q}\left( x \right) \approx {r_{Q(0)}}+ qx\sin \left( {{\theta _{Q(0)}}} \right)$ where $q=1$ if $Q=BR$ and $q=-1$ if $Q=RU$.
\vspace{-0.2cm}
\subsection{Path Loss Model}

Since surfaces are always installed on the walls of buildings, adjusting angles of surface physically may have constraints based on the shapes or directions of the walls. In this scenario, we operate the surfaces as anomalous reflectors, which is configured for reflecting waves towards a distinct direction of users. Hence, waves can be emitted to various directions with unequal angles of incidence and reflection.

\begin{lemma}\label{lemma2}
We assume surfaces act as RISs, thereby obtaining the coefficient $\Theta \left( x \right) = \left( {\sin \left( {{\theta _{BR\left( 0 \right)}}} \right) - \sin \left( {{\theta _{RU\left( 0 \right)}}} \right)} \right)x + {{{\phi _0}}}/{k}$. Considered on the long-distance regions with directional lights, the pass loss model on RIS operations can be derived as
\begin{align}\label{PtRIS}
P_t^{RIS} \approx {C ^2_{RIS}}{\left( {{r_{{B}R\left( 0 \right)}}{r_{{R_i}U\left( 0 \right)}}} \right)^{ - {{{\alpha _t}}}}},
\end{align}
where ${C _{RIS}} = \frac{L}{{4\pi }}\left( {\cos \left( {{\theta _{{B}R\left( 0 \right)}}} \right) + \cos \left( {{\theta _{RU\left( 0 \right)}}} \right)} \right)$ and $\alpha_t$ is the path loss exponent of typical users.
\begin{IEEEproof}
Substituting $\Theta \left( x \right)$ into $\Psi\left( x \right)$, it is simplified as $\Psi\left( x \right) ={r_{{B}R\left( 0 \right)}} + {r_{RU\left( 0 \right)}} - \frac{{{\phi _0}}}{k}$. Since the BSs are assumed as point sources emitting cylindrical waves, we have assumptions such as ${r_Q}\left( x \right) \approx {r_{Q(0)}}+ qx\sin \left( {{\theta _{Q(0)}}} \right)$. Thus, the pass loss model can be approximated as
\begin{align}
P_t^{RIS} &\approx {\left| {\frac{L}{{{\rm{4}}\pi }}{{\left( {{r_{{B}R}}\left( x \right){r_{RU}}\left( x \right)} \right)}^{ - \frac{{{\alpha _t}}}{2}}}} \right|^2}\notag\\
&\hspace*{0.3cm} \times {\left| {\left( {\cos \left( {{\theta _{{B}R\left( 0 \right)}}} \right) + \cos \left( {{\theta _{RU\left( 0 \right)}}} \right)} \right)} \right|^2} \notag \\
 &\hspace*{0.3cm} \times {\left| {\exp \left( { - jk\left( {{r_{{B}R\left( 0 \right)}} + {r_{RU\left( 0 \right)}} - \frac{{{\phi _0}}}{k}} \right)} \right)} \right|^2},
\end{align}
and we obtain \eqref{PtRIS} via algebraic manipulations.
\end{IEEEproof}
\end{lemma}

\subsection{Distance Distributions}

Note that the users and BSs are settled via two independent HPPPs and reflecting surface are uniformly deployed within the ball $\mathbb{O}(0,R_L)$ of typical users. The emphasis quantities of this network are the distance distributions of three links, such as BU, BR, and RU links. Based on the MCP pattern of PCP models, locations are defined that BSs and users are parent nodes obeying HPPPs and surfaces are daughter nodes within the clusters of RIS balls. Via the aforementioned settings, we derive the probability density functions (PDFs) of distances of the corresponding cluster and other clusters for a typical user.

\subsubsection{The Corresponding Cluster of the Typical User}
We focus on a typical user located at the center of the RIS ball area served by uniformly distributed intelligent surfaces. Thus, we can derive the PDF of the distance from a surface to its targeted typical user, denoted as $r_{RU}$, as
\begin{align}\label{PDFRU}
{f_{{r_{RU}}}}\left( x \right) = \frac{{2x}}{{R_L^2}}U\left( {{R_L} - x} \right),
\end{align}
where $U\left( \cdot \right)$ is the unit step function.

\subsubsection{Other Clusters of the Typical User}
Since the LOS links from BSs to the typical user are blocked, we can investigate the reflecting links from the BSs to the surfaces. Thus, based on the null probability of a 2-D PPP within in the RIS ball area and order statistics \cite{nNearestUser}, the PDF of the distance between a RIS and its $n^{th}$ nearest BS is derived as
\begin{align}\label{r_BU}
{f_{{r_{BR}}}}\left( {x,n} \right) = \frac{{2{{\left( {\pi {\lambda _b}} \right)}^n}}}{{\left( {n - 1} \right)!}}{x^{2n - 1}}\exp \left( { - \pi {\lambda _b}{x^2}} \right).
\end{align}

\subsection{Angle Distributions}

Shown as Fig. \ref{fig_1}, we denote a BS as node $B$, a RIS as node $R$, and a typical user as node $U$ to clarify the angles. With the aid of a chosen positive X-axis that is parallel to the RIS, the angles are illustrated as $\psi_1 = \angle{RUX}  $, $\psi_2 = \angle{BRX}$ and $\theta = \left|\pi-\left|\psi_2-\psi_1\right|\right|$. Notice that the angle of $\psi_2$ is uniformly distributed within $(0,2\pi)$ based on the properties of HPPP. We additionally observe that the angle of $\psi_1$ obeys uniform distribution in $(0,2\pi)$ since the RIS is uniformly distributed in the RIS ball. Based on $\psi_1$ and $\psi_2$ with the same distributions, the cumulative distribution function (CDF) of $ \left|\psi_2-\psi_1\right|$ is derived as
\begin{align}
{F_{\left| {{\psi _{\rm{2}}} - {\psi _1}} \right|}}\left( z \right)  &= \frac{{4\pi z - {z^2}}}{{4{\pi ^2}}},
\end{align}
therefore, the PDF of the angle of $ \left|\psi_2-\psi_1\right|$ can be derived as ${f_{\left| {{\psi _{\rm{2}}} - {\psi _1}} \right|}}\left( z \right) = \frac{{2\pi  - z}}{{2{\pi ^2}}}$.

With the respect to $\theta = \left|\pi-\left|\psi_2-\psi_1\right|\right|$, the CDF of the angle $\theta$ is derived as
\begin{align}
&{F_\theta }\left( x \right) = {F_{\left| {{\psi _{\rm{2}}} - {\psi _1}} \right|}}\left( {x + \pi } \right) - {F_{\left| {{\psi _{\rm{2}}} - {\psi _1}} \right|}}\left( {x - \pi } \right) = \frac{x}{\pi },
\end{align}
which is proved that the angle $\theta$ obeys uniform distribution within $(0,\pi)$ with the PDF as $f_\theta (x) = 1/\pi$.

Recall that we denote the angles of incidence as $\theta _{BR\left( 0 \right)}$ and the angles of reflection as $\theta _{RU\left( 0 \right)}$, thereby we can observe $\theta = \theta _{BR\left( 0 \right)}+\theta _{RU\left( 0 \right)}$ from Fig. \ref{fig_1}. In the following, the angle analysis with respect to RISs is investigated.

When the surfaces are designed as RISs, the angles of incidence and reflection are unequal. We define the angles of incidence $\theta _{BR\left( 0 \right)}=\rho_a \theta$, where $\rho_a \in(0,1)$, thereby the angles of reflection is $\theta _{RU\left( 0 \right)} = (1-\rho_a)\theta$. Hence, the PDFs of the angles of incidence and reflection can be derived as
\begin{align}
&\label{angle1}{f_{{\theta _{BR\left( 0 \right)}}}}(x) =  \frac{1}{{{\rho _a}\pi }},x \in \left( {0,\frac{\pi }{2}} \right),\\
&\label{angle2}{f_{{\theta _{RU\left( 0 \right)}}}}(x) = \frac{1}{{\left( {1 - {\rho _a}} \right)\pi }},x \in \left( {0,\frac{\pi }{2}} \right).
\end{align}

\begin{lemma}\label{EverageAngle}
Considered on an average case with a 2-D HPPP, the angles are uniformly distributed. Considered random spacial effect, the average path loss of RIS links with respect to angles can be derived as
\begin{align}
P_t^{RIS} = {C _{RIS,E}}{\left( {{r_{{B}R\left( 0 \right)}}{r_{{R_i}U\left( 0 \right)}}} \right)^{ - {{{\alpha _t}}}}},
\end{align}
\emph{where ${C _{RIS,E}}  =\frac{{{L^2}}}{{16{\pi ^3}}}\left( {\pi  + \frac{{\sin \left( {2{\rho _a}\pi } \right)}}{{4{\rho _a} - 12\rho _a^2 + \rho _a^3}}} \right)$.}
\begin{IEEEproof}
The average path loss with respect to angles can be expressed as $P_t^{RIS} = \mathbb{E}\left[ {C_{RIS,I}^2} \right]{\left( {{r_{{B}R\left( 0 \right)}}{r_{{R_i}U\left( 0 \right)}}} \right)^{ - {{{\alpha _t}}}}}$ and via plugging \eqref{angle1} and \eqref{angle2}, the lemma can be proved.
\end{IEEEproof}
\end{lemma}

\section{Performance Evaluation}

Via pre-deciding a fixed threshold rate, the communication performance can be guaranteed when the transmit rates are higher than the threshold. By this method to evaluate whether the QoS of a network is satisfied, we investigate the SINR coverage performance on our RIS-aided NOMA networks based on path loss models of reflecting links.

The coverage probability expressions for a connected user and a typical user can be expressed respectively as
\begin{align}
&\label{coverage_connected}{\mathbb{P}}_{t}={\rm{Pr}}\left\{ {{\gamma _{{\rm{SIC}}}} > \gamma _{SIC}^{th},{\gamma _t} > \gamma _t^{th},\gamma _t^{OMA} >\mathbb{E}\left[ {\gamma _c^{{OMA}}} \right]} \right\},\\
&\label{coverage_typical}\mathbb{P}_{c}={\rm{Pr}}\left\{ {{\gamma _{\rm{c}}} > \gamma _c^{th},\mathbb{E}\left[ {\gamma _t^{OMA}} \right] > \gamma _c^{OMA}} \right\},
\end{align}
where $\mathbb{P}$ is the probability, $\mathbb{E}\left[\cdot\right]$ is the mean value, $\gamma _t^{OMA} = \frac{{{P_b}{{\left| {{h_{t,{{\bf{x}}_B}}}} \right|}^2}P_t^{RIS}}}{{{I_{t,{\rho _t}}} + {\sigma ^2}}}$ and $\gamma _c^{OMA} = \frac{{{P_b}{{\left| {{h_{c,{{\bf{x}}_B}}}} \right|}^2}C{r_c}^{ - {\alpha _c}}}}{{{I_c} + {\sigma ^2}}}$ are the received signal-to-noise-ratio (SNR) in OMA scenarios, which determine the SIC order, ${\gamma _{SIC}^{th}}$ is the threshold of SIC procedures, $\gamma _t^{th} = {2^{{R_t}/{B_w}}} - 1$ is the coverage threshold of the typical user with threshold rate $R_t$ and bandwidth $B_w$, $\gamma _c^{th} = {2^{{R_c}/{B_w}}} - 1$ is the threshold of the connected user with threshold rate $R_c$.
\vspace{-0.2cm}
\subsection{Interference Analysis}
Before evaluating the coverage performance of this network, we would first derive the Laplace transform of interference, ${I_{t,{\rho _t}}}$ and $I_{c}$, under two scenarios for tractability of the analysis.

\subsubsection{Interference Analysis of a Connected User}
Since the connected user is not served by reflecting surfaces, the Laplace transform of the interference for a connected user can be expressed via traditional wireless communication analysis as
\begin{align}
&{\cal L}_{(s)}^{c} =\mathbb{E} \left[ \exp\left(-{\sum\limits_{{{\bf{x}}_I} \in {\Phi _r} \setminus {{\bf{x}}_B}} {{P_b}C{{\left| {{h_{t,{{\bf{x}}_I}}}} \right|}^2}{r_{c,I}}^{ - {\alpha _c}}} }\right) \right].
\end{align}

\begin{lemma}\label{I_c}
The Laplace transform of interference for the connected user can be derived as
\begin{align}
{\cal L}_{(s)}^c = \exp \left( { - {\varsigma _1}\left( {{}_2{F_1}\left( { - \frac{2}{{{\alpha _c}}},m;1 - \frac{2}{{{\alpha _c}}}; - \varsigma_2 s} \right) - 1} \right)} \right),
\end{align}
where ${}_2{F_1}\left( \cdot,\cdot;\cdot; \cdot \right)$ is the hypergeometric function, ${\varsigma _1} = \pi {\lambda _b}r_c^2$ and ${\varsigma _2} = \frac{{{P_b}C}}{{mr_c^\alpha }}$.
\begin{IEEEproof}
Considering Campbell's theorem and probability generating functional (PGFL), this lemma can be proved.
\end{IEEEproof}
\end{lemma}

\subsubsection{Interference Analysis of a Typical on RIS Scenario}
With the aid of \textbf{Lemma \ref{lemma2}}, the Laplace transform of the interference under RIS scenarios can be expressed as
\begin{align}
&{\cal L}_{(s)}^{t,RIS} = \mathbb{E}\left[ \exp\left(-{{\rho _t}\sum\limits_{{{\bf{x}}_I} \in {\Phi _r} \setminus {{\bf{x}}_B}} {\frac{{{P_b}C_{RIS,I }^2{{\left| {{h_{t,{{\bf{x}}_I}}}} \right|}^2}}}{{{{\left( {{r_{BR\left( I  \right)}}{r_{RU\left( I  \right)}}} \right)}^{{\alpha _t}}}}}} } \right)\right],
\end{align}

\begin{lemma}\label{I_t}
With the aid of RISs, the Laplace transform of interference for the typical users can be derived as
\begin{align}
 &{\cal L}_{(s)}^{t,RIS}\left( {{r_{BR\left( 0 \right)}},{r_{RU\left( 0 \right)}}} \right)=\notag\\
 &\hspace*{0.3cm} \exp \left( { - {\varsigma _3}\left( {{}_2{F_1}\left( { - \frac{2}{{{\alpha _t}}},m;1 - \frac{2}{{{\alpha _t}}}; - {{s{\varsigma _4}}}{{}}} \right) - 1} \right)} \right),
\end{align}
where ${\varsigma _3} = \pi {\lambda _b}r_{BR\left( 0 \right)}^2$ and ${\varsigma _4} = {P_b}C_{RIS,E}^2\frac{{1}}{{{m_t}{r_{RU\left( 0 \right)}}{r_{BR\left( 0 \right)}}^{{\alpha _t}}}}$.
\begin{IEEEproof}
Same as \textbf{Lemma \ref{I_c}}.
\end{IEEEproof}
\end{lemma}

\subsection{Coverage Analysis with RISs}
In this subsection, the coverage probabilities of the typical users and the connected users are derived with the aid of RISs.

\subsubsection{Coverage Analysis of the Typical User with RISs}
Note that the interference from the typical user is strived to be canceled with the aid of SIC procedures. When the surfaces perform as RISs, based on \textbf{Lemma \ref{lemma2}} and \eqref{coverage_typical}, the coverage probability can be rewritten as
\begin{align}
{\mathbb{P}}_{t} = {\rm{Pr}}\left\{ {{{\left| {{h_{t,{{\bf{x}}_B}}}} \right|}^2} > \frac{{\Upsilon \left( {{I_{t,{\rho _t}}} + {\sigma ^2}} \right)}}{{{P_b}P_t^{RIS}}}} \right\},
\end{align}
where $\Upsilon  = \max \left( {\frac{{\gamma _{SIC}^{th}}}{{{a_c} - \gamma _{SIC}^{th}{a_t}}},\frac{{\gamma _t^{th}}}{{{a_t}}},\gamma _c^E} \right)$ and $\gamma _c^E = \mathbb{E}\left[ {\frac{{{P_b}{{\left| {{h_{c,{{\bf{x}}_B}}}} \right|}^2}C{r_c}^{ - {\alpha _c}}}}{{{I_c} + {\sigma ^2}}}} \right]$.

\begin{theorem}\label{CPRIS}
Note that we assume reflecting channels as Nakagami-m fading channels. With the aid of RISs, the approximated expressions of coverage probability for the typical users can be derived as
\begin{align}\label{CPRRISexact}
{\mathbb{P}}_{t} &\approx  2\pi {\lambda _b}\int_0^{{R_L}} {\int_0^\infty  {\sum\limits_{n = 1}^{{m_t}} {{{\left( { - 1} \right)}^{n + 1}} {\binom{m_t}{n}} } x} }\notag \\
&\hspace*{0.3cm}\times\exp \left( { - {\beta _{\rm{0}}}\left( y \right){x^{{\alpha _t}}}} \right)\exp \left( { - {\beta _2}{x^2}} \right)dx{f_{{r_{RU}}}}\left( y \right)dy,
\end{align}
where ${\beta _{\rm{0}}}\left( x \right){\rm{ = }}{\beta _{\rm{1}}}x^\alpha_t$, ${\beta _{\rm{1}}}{\rm{ = }}\frac{{n{\eta _t}\Upsilon {\sigma ^2}}}{{{P_b}C_{RIS,E}^2}}$, ${\beta _2} = \pi {\lambda _b}{}_2{F_1}\left( { - \frac{2}{{{\alpha _t}}},m;1 - \frac{2}{{{\alpha _t}}}; - \frac{{n{\eta _t}\Upsilon }}{{{m_t}}}} \right)$ and $m_t$ is the coefficient in Nakagami-m fading channels with unit mean values.
\begin{IEEEproof}
See Appendix~A.
\end{IEEEproof}
\end{theorem}

\begin{corollary}\label{CPRIS2}
Conditioned on $\alpha_t = 4$, we can derive the closed-form expressions of the coverage probability for the typical user as
\begin{align}
{\mathbb{P}}_{t} &\approx  \sum\limits_{n = 1}^{{m_t}} {{{\left( { - 1} \right)}^{n + 1}}{\binom{m_t}{n}} } \sum\limits_{i = 1}^K {\frac{{{\omega _i}{\pi ^{\frac{3}{2}}}{\lambda _b}\sqrt {1 - \Xi _i^2} }}{{2R_L^{}\sqrt {{\beta _1}} {\Xi _i}}}} \notag\\
 &\hspace*{0.3cm}\times \exp \left( {\frac{{\beta _2^2}}{{4{\beta _1}\Xi _i^4}}} \right)\rm{Erfc}\left( {\frac{{\beta _2^{}}}{{2\sqrt {{\beta _1}} \Xi _i^2}}} \right),
\end{align}
where ${\eta _t} = {m_t}{\left( {{m_t}!} \right)^{ - \frac{1}{{{m_t}}}}}$, ${\varpi _i}{\rm{ = cos}}\left( {\frac{{2i - 1}}{{2K}}\pi } \right)$, ${\Xi _i} = \frac{{{R_L}}}{2}\left( {{\varpi _i} + 1} \right)$, ${\omega _i} = \pi /K$ and $\rm{Erfc}(\cdot)$ is the error function.
\begin{IEEEproof}
Based on Appendix~A when $\alpha_t=4$, this corollary can be proved via substituting \eqref{PDFRU} and Chebyshev-gauss quadrature into \eqref{CPRRISexact}.
\end{IEEEproof}
\end{corollary}

\subsubsection{Coverage Analysis of the Connected User}
Based on \eqref{coverage_connected}, we can rewrite the coverage probability expressions as
\begin{align}
{\mathbb{P}}_{c} =  &{\rm{Pr}}\left\{ {{{\left| {{h_{c,{{\bf{x}}_B}}}} \right|}^2} > \frac{{\gamma _t^E\left( {{I_c} + {\sigma ^2}} \right)}}{{P_bC{r_c}^{ - {\alpha _c}}}}} \right\} \notag\\
&- {\rm{Pr}}\left\{ {{{\left| {{h_{c,{{\bf{x}}_B}}}} \right|}^2} > \frac{{\gamma _c^{th}\left( {{I_c} + {\sigma ^2}} \right)}}{{\left( {{a_c} - {a_t}\gamma _c^{th}} \right){P_b}C{r_c}^{ - {\alpha _c}}}}} \right\},
\end{align}
where $\gamma _t^E =\mathbb{E}\left[ P_b\frac{{{{\left| {{h_{t,{{\bf{x}}_B}}}} \right|}^2}P_t^{RIS}}}{{{I_{t,{\rho _t}}} + {\sigma ^2}}}\right]$.

\begin{theorem}\label{CPC}
The closed-form expressions of coverage probability for the connected users can be derived as
\begin{align}\label{CPCexact}
{\mathbb{P}}_c &\approx  \sum\limits_{n = 1}^{{m_c}} {{{\left( { - 1} \right)}^{n + 1}} {\binom{m_c}{n}} } \exp \left( { - {\mu _1}r_c^2 - {\mu _2}{r_c}^{{\alpha _c}}} \right)\notag\\
 &\hspace*{0.3cm}- \sum\limits_{n = 1}^{{m_c}} {{{\left( { - 1} \right)}^{n + 1}} {
\binom{m_c}{n}
}} \exp \left( { - {\mu _3}r_c^2 - {\mu _4}{r_c}^{{\alpha _c}}} \right),
\end{align}
where ${\mu _1} = \pi {\lambda _b}\left( {{}_2{F_1}\left( { - \frac{2}{{{\alpha _c}}},m;1 - \frac{2}{{{\alpha _c}}}; - \frac{{n{\eta _c}{P_b}\gamma _t^E}}{m}} \right) - 1} \right)$, ${\mu _3} = \pi {\lambda _b}\left( {{}_2{F_1}\left( { - \frac{2}{{{\alpha _c}}},m;1 - \frac{2}{{{\alpha _c}}}; - \frac{{n{\eta _c}\gamma _c^{th}}}{{m\left( {{a_c} - {a_t}\gamma _c^{th}} \right)}}} \right) - 1} \right)$, ${\mu _2} = {C^{ - 1}}n{\eta _c}\gamma _t^E{\sigma ^2}$, ${\mu _4} = \frac{{n{\eta _c}\gamma _c^{th}{\sigma ^2}}}{{\left( {{a_c} - {a_t}\gamma _c^{th}} \right){P_b}C}}$ and ${\eta _c} = {m_c}{\left( {{m_c}!} \right)^{ - \frac{1}{{{m_c}}}}}$ with Nakagami-m fading coefficient $m_c$.
\begin{IEEEproof}
Same as Appendix~A with $\gamma _t^E \approx E\left[ {\frac{{\left( {{\alpha _t} - 2} \right)}}{{\pi {\lambda _b}r_{BR\left( 0 \right)}^2}}} \right] \approx c_t^E\left( {{\alpha _t} - 2} \right)\pi {\lambda _b}$, where $c_t^E = {10^{ - 2}} \times \Gamma \left( {{{10}^{ - 2}}} \right)$ based on Eq. [2.3.18.2] in \cite{table}.
\end{IEEEproof}
\end{theorem}

\section{Numerical Results}

In this section, numerical results validate analytical coverage probability for typical users (\textbf{Theorem \ref{CPRIS}}) and connected users (\textbf{Theorem \ref{CPC}}) as upper or lower bounds. Without otherwise specification, we define the numerical settings as: the half-length of RISs as $0.75$ m, the noise power as ${\sigma ^2} =  - 170 + 10\log \left( {f_c} \right) + {N_f}=-90 $ dB, where $f_c$ is the bandwidth as $10$ MHz and $N_f$ is noise figure as $10$ dB, transmit power of users as $\left[0,15\right]$ dBm, pass loss exponents as $\alpha_c = 4$ and $\alpha_t = 2.4$, RIS ball radius $R_L = 25$ m, density of BSs as $\lambda_b = 1/(300^2\pi)$, thresholds $\gamma_{SIC}^{th}=\gamma_{t}^{th}=\gamma_c^{th}=10^{-2}$, Gamma distribution coefficient $m_c=m_t=4$ and power allocation coefficients $a_c = 0.6$ and $ a_t = 0.4$.

\begin{figure}[!htb]
\vspace{-0.2cm}
\centering
\includegraphics[width= 2.5in]{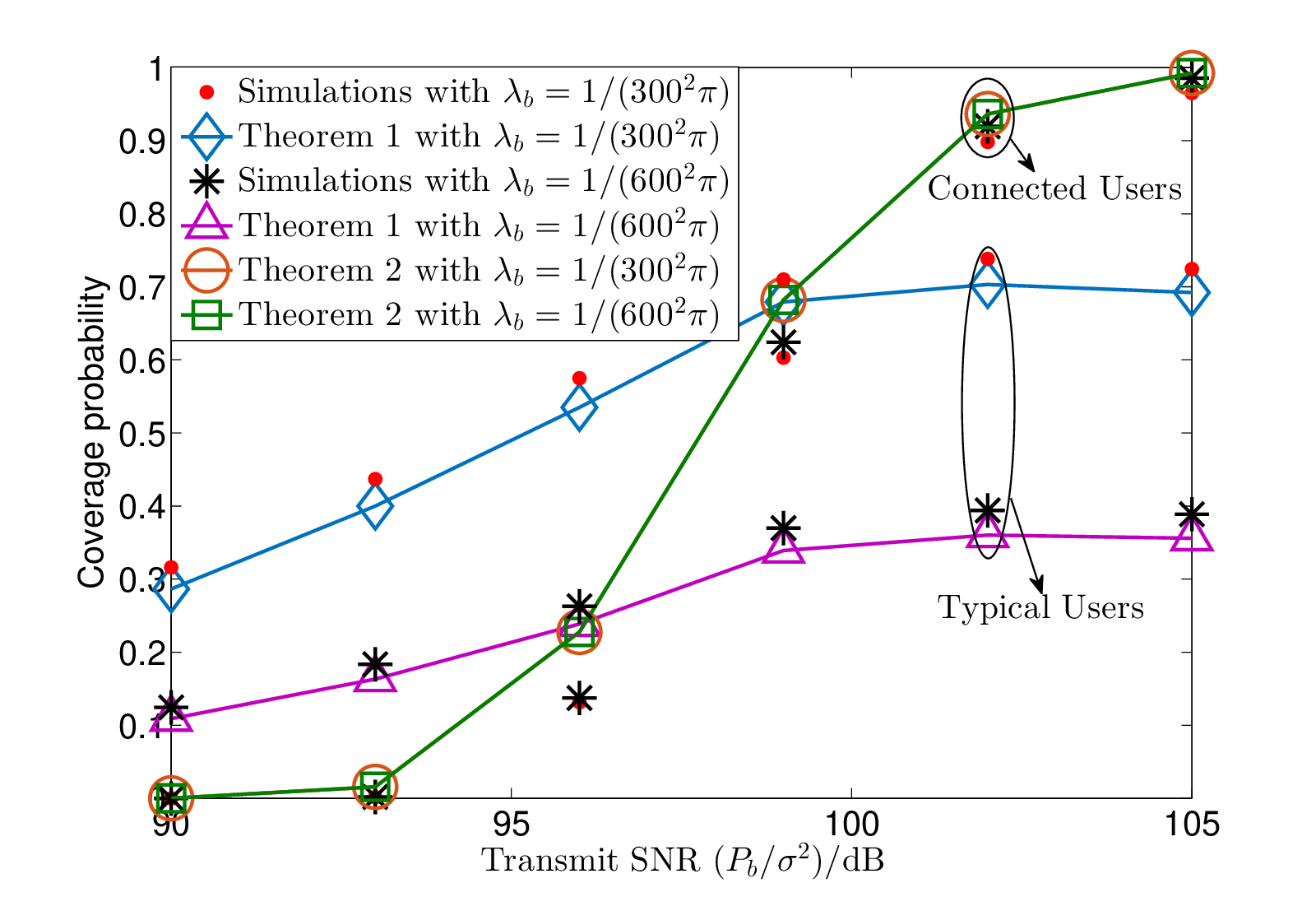}
\vspace{-0.2cm}
\caption{Coverage probability versus transmit SNR with various density of BSs $\lambda_b=[1/(300^2\pi),1/(600^2\pi)]$: a verification of \textbf{Theorem \ref{CPRIS}} and \textbf{Theorem \ref{CPC}}. }
\label{figure1}
\vspace{-0.3cm}
\end{figure}

Fig. \ref{figure1} illustrates that the analytical coverage probability expressions of typical users (\textbf{Theorem \ref{CPRIS}}) perform as upper bounds, which is because we exploit the upper bounds of normalized Gamma variables. In addition, the analytical expressions for connected users (\textbf{Theorem \ref{CPC}}) are observed as lower bounds since the evaluated average channel gain threshold $\gamma_t^E$ is higher than the exact value, which provokes low probabilities of the SIC condition $\gamma_c^{OMA}<\gamma_t^E$. Another illustration is that the density of BSs significantly affects the coverage performance of typical users while it has a slight influence on connected users. This is because the total interference of the typical users is more sensitive than that of the connected users, thereby enhancing the density of BSs enlarges the interference of typical users much more severely than the connected users.

\begin{figure}[!htb]
\vspace{-0.2cm}
\centering
\includegraphics[width= 2.5in]{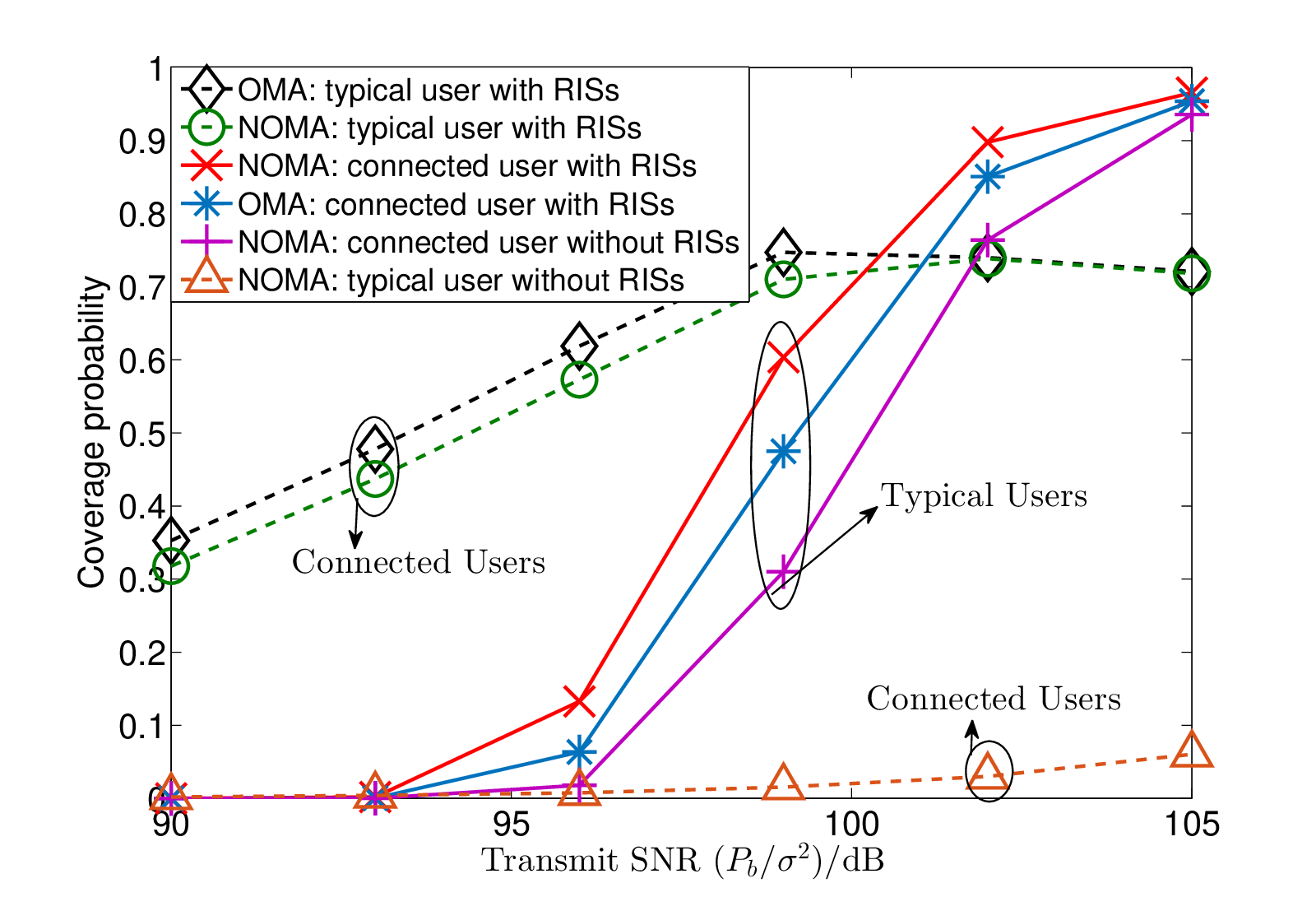}
\vspace{-0.2cm}
\caption{Coverage probability versus transmit SNR: a comparison among traditional NOMA, RIS-aided OMA and RIS-aided NOMA scnearios.}
\label{figure2}
\vspace{-0.3cm}
\end{figure}
\begin{figure}[!htb]
\vspace{-0.2cm}
\centering
\includegraphics[width= 2.5in]{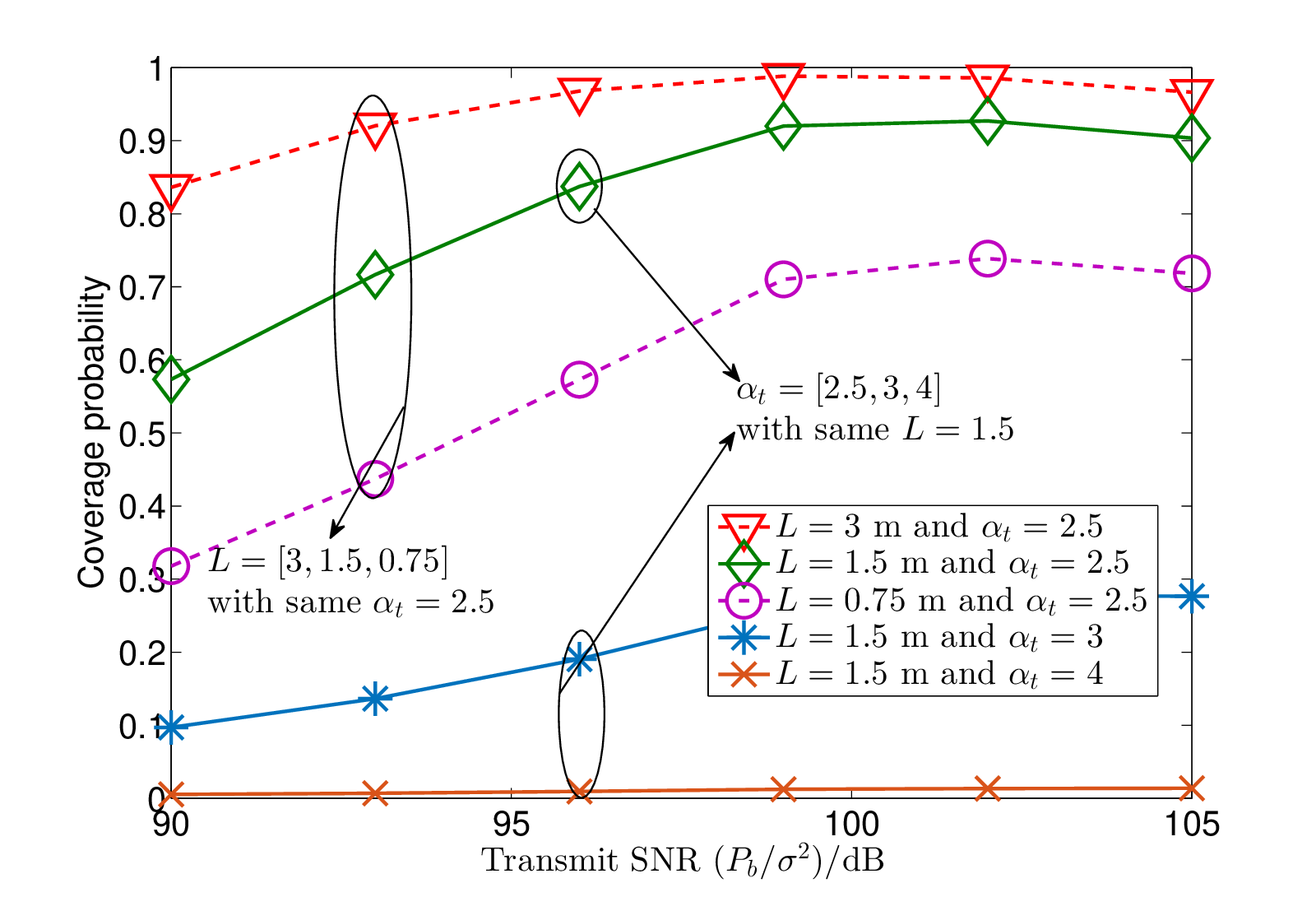}
\vspace{-0.2cm}
\caption{Coverage probability versus transmit SNR with various lengths of RISs $L = [0.75, 1.5 ,3]$ m and path loss exponents $\alpha_t=[2.5,3,4]$.}
\label{figure3}
\vspace{-0.3cm}
\end{figure}

The performance among traditional NOMA, RIS-aided NOMA, and RIS-aided OMA scenarios are compared in Fig. \ref{figure2}, which demonstrates that the performance of NOMA users boosts considerably with the aid of RISs, especially for typical users. The enhancement of NOMA users can be explained that 1) when assisted with RISs, the connected users can avoid SIC procedures since the typical users with substantially improved channel gains take over the SIC procedures, thereby connected users would not experience outage scenarios caused by SIC failures; 2) with the aid of RISs, superior channel gains of typical users increase coverage performance.

Considering the length of RISs, Fig. \ref{figure3} investigates the performance varied by $L$ and path loss exponents. Two observations are apparent to explain that: 1) long lengths of RISs cause high performance since more reconfigurable meta-material elements are involved and 2) enlarging path loss exponents results in reduced performance as the relationship between the path loss exponents and the coverage performance is a negative correlation.

\section{Conclusion}

This paper has investigated the coverage probability of RIS-assisted NOMA frameworks, where the PCP principle is invoked to capture the spatial effects of NOMA users. The path loss models of RIS reflecting links has been derived, which is correlated with the product of distances to conform with long-distance regions. The angle distributions have been presented with a conclusion that the BS-RIS-user angles obey uniform distributions in $[0.\pi]$. With the aid of the derived closed-form expressions of coverage probabilities and numerical results, the performance of traditional NOMA, RIS-aided NOMA, and RIS-aided OMA scenarios have been compared, which concluded that RISs enhance the performance of users significantly. The analysis of this paper has verified that two applications of RISs in multi-cell NOMA networks are feasible, such that: 1) RISs can improve the channel conditions of blocked or far users; 2) RISs can alter the SIC order to maintain primary users avoiding SIC procedures.

\vspace{-0.2cm}
\section*{Appendix~A: Proof of Theorem~\ref{CPRIS}} \label{Appendix:A}
\renewcommand{\theequation}{A.\arabic{equation}}
\setcounter{equation}{0}
\bibliographystyle{IEEEtran}

Based on PGFL, the expressions of the average interference for the connected user can be derived as
\begin{align}
\mathbb{E}\left[ {{I_c}} \right] = 2\pi {\lambda _b}\int_{{r_c}}^\infty  {{P_b}C{r^{1 - {\alpha _c}}}} dr{\rm{ = }}\frac{{2\pi {\lambda _b}{P_b}Cr_c^{2 - {\alpha _c}}}}{{{\alpha _c} - 2}}.
\end{align}
and via $\mathbb{E}\left[ {{I_c}} \right]$, $\gamma_c^E$ can be derived as
\begin{align}
\gamma _c^E = \frac{{\left( {{\alpha _c} - 2} \right){P_b}C{r_c}^{ - {\alpha _c}}}}{{2\pi {\lambda _b}{P_b}Cr_c^{2 - {\alpha _c}} + \left( {{\alpha _c} - 2} \right){\sigma ^2}}}.
\end{align}

Note that the normalized Gamma variables have a tight upper bound, denoted as $\mathbb{P}\left[\left|h^2\right|<x\right] <\left(1-e^{-x{\eta _t}}\right)^{m_t}$. Utilizing binomial expansions, the expressions of coverage probability for the typical user can be expressed as
\begin{align}
{\mathbb{P}}_{t} {\approx } \sum\limits_{n = 1}^{{m_t}} {{{\left( { - 1} \right)}^{n + 1}}{\binom{m_t}{n}}} {\cal L}\left[ {\frac{{n{\eta _t}\Upsilon {I_{t,{\rho _t}}}}}{{{P_b}P_t^{RIS}}}} \right]\mathbb{E}\left[ {{e^{ - \frac{{n{\eta _t}\Upsilon {\sigma ^2}}}{{{P_b}P_t^{RIS}}}}}} \right],
\end{align}
and via substituting \textbf{Lemma \ref{EverageAngle}} and \textbf{Lemma \ref{I_t}} into the equation above, the theorem can be verified.

\bibliographystyle{IEEEtran}
\bibliography{mybib}
\end{document}